\documentclass{article}

\newtheorem{thm}{Theorem}

\title {A Simple In-Place Algorithm for In-Shuffle.}
\author {Peiyush Jain, Microsoft Corporation.}
\date {July 2004}

\begin{document}
\maketitle

\section{Introduction}
\label{}

An in-shuffle of a deck of cards is done by cutting the deck into two equal
halves and interleaving them perfectly, with the first card of the second half
being the first card of the shuffled deck. In case of \( 2n \) cards, this
kind of a shuffle gives rise to a permutation (called the \emph{in-shuffle permutation
of order \( 2n \)}) given by:
\begin{displaymath}
i \rightarrow 2i\mbox{ mod } (2n+1), i  \in  \{1, 2,\ldots, 2n\}
\end{displaymath}
In \cite{EM00} it has been shown that the problem of merging lists in-place
can be reduced to performing in-shuffles in-place. In the same paper, the authors
provide a linear time, in-place algorithm for performing the in-shuffle, when
the list is represented as an array, indexed \( 1 \) to \( 2n \).
In this paper we provide a simpler algorithm which permits an easy implementation.
\section{Preliminaries}
Any permutation is a set of disjoint cycles. For e.g. the in-shuffle of 6 elements
is composed of two cycles: \( (1,2,4) \) and \( (3,6,5) \). It is easy to move the elements
of a single cycle using just an extra location, by a {}``cycle leader{}''
algorithm, \cite{FMP95}. This algorithm proceeds by repeatedly making a space
in the list, computing the index of element that belongs to that space and moving
that element, so creating a new space.

For any permutation, we can proceed by applying the cycle leader algorithm to
each cycle in turn in order to realise the whole permutation. If, for a permutation,
we can compute the location of an element of the 'next' cycle in constant time and minimal space, we could realise
the whole permutation using a linear time and an in-place algorithm.

We show that, when \( 2n \) is of the form \( 3^{k} \)\( -1 \), we can easily
determine the cycles of the in-shuffle permutation of order \( 2n \). We will need the following theorem
from number theory:

\begin{thm} If \( p \) is an odd prime and \( g \) is a primitive root of \( p^{2} \), then \( g \) 
is a primitive root of \( p^{k} \) for any \( k \)\( \geq 1 \).
\end{thm}
A proof of this theorem can be found in \cite[p 20-21]{Nar00}.

It can be easily seen that \( 2 \) is a primitive root of \( 9 \). From the
above theorem it follows that \( 2 \) is also a primitive root of \( 3^{k} \)
for any \( k \)\( \geq 1 \). This implies that the group \( (Z/3^{k})^{*} \)
is cyclic with \( 2 \) being its generator.

Now let us analyse the cycles of an in-shuffle permutation when \( 2n \) \( = \)
\( 3^{k}-1 \).

The cycle containing \( 1 \) is nothing but the group \( (Z/3^{k})^{*} \), which
consists of all numbers relatively prime to \( 3^{k} \) and less than it. 

Let \( 1 \leq s < k \).  Consider the cycle containing \( 3^{s} \). Every number in this cycle is of
the form \( 3^{s} 2^{t}\ (modulo\  3^{k}) \) for \( 1\leq t\leq \varphi (3^{k}) \) (where $\varphi$ is the Euler-totient function).
Since 2 is a generator of \( (Z/3^{k})^{*} \), this cycle contains \emph{exactly} the
numbers less than \( 3^{k} \) which are divisible by \( 3^{s} \) but not by
 any higher power of \( 3 \).

This means that in an in-shuffle permutation of order \( 3^{k}-1 \), we have
exactly \( k \) cycles with \( 1,3,3^{2},\ldots ,3^{k-1} \) each belonging to
a different cycle. Thus for these permutations, it becomes easy to pick the 'next'
cycle in order to apply the cycle leader algorithm. Note that the length of the cycle
containing $3^s$ is $\varphi(3^k)/3^s$, which helps us implement the cycle leader algorithm 
more efficiently.

We present the algorithm for the general case of an in-shuffle permutation
of order \( 2n \) in the next section.
\section{Algorithm}
We start with an observation. Suppose for each \( 2n \) we can easily find
a \( 2m \) (= \( \Omega (n) \) and \( \leq 2n \)) for which we have a linear
time, in-place, in-shuffle algorithm, we then have a linear time, in-place, in-shuffle
algorithm for \( 2n \). This is because we can shuffle the array
\begin{displaymath}
[a_{1}, a_{2}, \ldots, a_{m}, a_{m+1}, \ldots, a_{n}, a_{n+1}, \ldots, a_{n+m}, a_{n+m+1}, \ldots, a_{2n}]
\end{displaymath}
to look like 
\begin{displaymath}
[a_{1}, a_{2}, \ldots, a_{m}, a_{n+1}, \ldots, a_{n+m}, a_{m+1}\ldots, a_{n}, a_{n+m+1}\ldots, a_{2n}]
\end{displaymath}
in linear time and in-place by doing a right cyclic shift of the elements \( a_{m+1},\ldots ,a_{n+m} \)
by a distance \( m \). We can achieve this by reversing the sub-array \([m+1, \ldots, n+m]\)  followed by a reversal of
the sub-arrays \([m+1, \ldots, n]\) and \([n+1, \ldots, n+m]\).
We can now perform an in-shuffle of the first \( 2m \) elements and recursively perform an in-shuffle of the remaining elements.

Thus we have the following {\bf{\em In-shuffle Algorithm:}}
\begin{description}
\item[Input:] An array \(A [1, \ldots, 2n] \)
\item[Step 1.] 
Find a $2m = 3^k-1$ such that $3^k \leq 2n < 3^{k+1}$
\item[Step 2.] 
Do a right cyclic shift of $A[m+1,\ldots, n+m]$ by a distance $m$
\item[Step 3.]
 For each $i \in \{0, 1, \ldots, k-1\}$, starting at $3^i$, do the cycle leader algorithm 
 for the in-shuffle permutation of order $2m$ 
\item[Step 4.] 
Recursively do the in-shuffle algorithm on $A[2m+1, \ldots, 2n]$.
\end{description}
\section{Space and Time Complexity}
Let $T(n)$ be the time taken on an input array of size $n$.
Let us look at the time and space bounds for each of the steps.
\begin{description}
\item[Step 1] can be done in-place in $O(\log n)$ time. 
\item[Step 2] can be done in-place in $O(n)$ time.
\item[Step 3] can be done in-place in $O(m)$ time. 
\item[Step 4] is a tail recursive call and hence can be implemented in constant space. 
The time taken by this call is $T(2(n-m))$.
\end{description}
Since $m = \Omega(n)$, it follows that the total time taken by this algorithm is $O(n)$.

Thus we have a linear time, in-place algorithm for the in-shuffle of a list of $2n$ elements.
\section{Conclusion}
We have described a linear time, in-place, in-shuffle algorithm which I believe is interesting, simple and easy to implement.
It is not obvious whether the approach used (\emph {using a $p$ such that $k$ is a primitive root of $p^2$}) in this
paper can be generalised for $k$-way shuffles, but it is probable that we can use this approach for fixed squarefree $k$,
\cite{HEA86}. For small $k$, this approach can definitely be used. 
 \section*{Acknowledgement} I  would like to thank Barukh Ziv for suggesting that I write up this paper. I would also like to thank  
Karthik Mahesh for pointing this problem to me and Anand Ganesh for the interesting discussions we had.

\end{document}